\newif\ifmulticol	\multicoltrue
\newif\ifshowgit	\showgittrue		
\newif\ifgitlocal	\gitlocalfalse		
\newif\ifbiblatex	\biblatextrue		
\newif\ifbibnum		\bibnumtrue 		
\newif\iflineno		\linenofalse
\newif\iftoc		\tocfalse

\newif\iflucida		\lucidafalse
\newif\ifcm			\cmfalse
\newif\iflibertine	\libertinefalse
\newif\ifcharter	\chartertrue

\newcommand*{\secnumdepth}{0}			
\newcommand*{\tocdepth}{2}				
\newcommand*{\reftitle}{References}		
\newcommand*{\mytitleskip}{-0.2in}		

\multicoltrue\showgittrue\gitlocaltrue\biblatexfalse\bibnumtrue

\newcommand*{\mydocfontsize}{\ifcharter11pt\else\iflibertine11pt\else10pt\fi\fi}
\newcommand*{\setcol}{\ifmulticol twocolumn\else onecolumn\fi}

\documentclass[\mydocfontsize,\setcol]{article}

\newcommand*{\pdfauthor}{Steven A.~Frank}
\newcommand*{\pdftitle}{Optimizing differential equations to fit data and predict outcomes} 

\input fitode.sty

\newcommand*{\bu}{\bmr{u}}
\newcommand*{\bS}{\bmr{S}}
\newcommand*{\bb}{\bmr{b}}

\newcommand*{\Ge}{\epsilon}

\usepackage{bm}
\newcommand{\bmr}[1]{\bm{\mathrm{#1}}}

\DeclarePairedDelimiter\abs{\lvert}{\rvert}
\DeclarePairedDelimiter\norm{\lVert}{\rVert}
\DeclarePairedDelimiter\angb{\langle}{\rangle}
\DeclarePairedDelimiter\lrb{\lbrack}{\rbrack}
\DeclarePairedDelimiter\lr{\lparen}{\rparen}
\DeclarePairedDelimiter\lrbr{\lbrace}{\rbrace}

\makeatletter
\let\oldabs\abs \def\abs{\@ifstar{\oldabs}{\oldabs*}}
\let\oldnorm\norm \def\norm{\@ifstar{\oldnorm}{\oldnorm*}}
\let\oldangb\angb \def\angb{\@ifstar{\oldangb}{\oldangb*}}
\let\oldlrb\lrb \def\lrb{\@ifstar{\oldlrb}{\oldlrb*}}
\let\oldlr\lr \def\lr{\@ifstar{\oldlr}{\oldlr*}}
\let\oldlrbr\lrbr \def\lrbr{\@ifstar{\oldlrbr}{\oldlrbr*}}
\makeatother

\newcommand*{\dd}{\textrm{d}}

\newcommand*{\Eq}[1]{eqn~\ref{eq:#1}}

\newcommand*{\dovr}[2]{\frac{\dd #1}{\dd #2}}

\newcommand*{\Figure}[1]{Figure~\ref{fig:#1}}
\newcommand*{\Fig}[1]{Fig.~\ref{fig:#1}}

\newcommand*{\boldrule}{\hrule height 1.2pt}
\newcommand*{\noterule}{\medskip\boldrule\medskip}	

\newcommand*{\mytitle}{\pdftitle}

\newcommand*{\myauthor}{Steven A.\ Frank}
\newcommand*{\myaddress}{Department of Ecology and Evolutionary Biology, University of California, Irvine, CA 92697--2525, USA}
\newcommand*{\mycontact}{\href{https://stevefrank.org}{https://stevefrank.org}}

\newcommand*{\mykeywords}{ecological dynamics, evolutionary dynamics, optimization, differential equation models, artificial neural networks, automatic differentiation, Julia programming language, Bayesian analysis, Langevin dynamics}
\newcommand*{\myrunninghead}{Fitting differential equations}

\setabstract{-0.07}{\iftoc-2.0\else1.22\fi}{%
Many scientific problems focus on observed patterns of change or on how to design a system to achieve particular dynamics. Those problems often require fitting differential equation models to target trajectories. Fitting such models can be difficult because each evaluation of the fit must calculate the distance between the model and target patterns at numerous points along a trajectory. The gradient of the fit with respect to the model parameters can be challenging. Recent technical advances in automatic differentiation through numerical differential equation solvers potentially change the fitting process into a relatively easy problem, opening up new possibilities to study dynamics. However, application of the new tools to real data may fail to achieve a good fit. This article illustrates how to overcome a variety of common challenges, using the classic ecological data for oscillations in hare and lynx populations. Models include simple ordinary differential equations (ODEs) and neural ordinary differential equations (NODEs), which use artificial neural networks to estimate the derivatives of differential equation systems. Comparing the fits obtained with ODEs versus NODEs, representing small and large parameter spaces, and changing the number of variable dimensions provide insight into the geometry of the observed and model trajectories. To analyze the quality of the models for predicting future observations, a Bayesian-inspired preconditioned stochastic gradient Langevin dynamics (pSGLD) calculation of the posterior distribution of predicted model trajectories clarifies the tendency for various models to underfit or overfit the data. Coupling fitted differential equation systems with pSGLD sampling provides a powerful way to study the properties of optimization surfaces, raising an analogy with mutation-selection dynamics on fitness landscapes.
}

\begin{document}

\mymaketitle

\iftoc\mytoc{-24pt}{\newpage}\fi

\section{Introduction}

Much of science describes or predicts how things change over time. Differential equations provide a common model for fitting data and predicting future observations. Optimizing a differential equation model is challenging. Each observed or predicted point along a temporal trajectory is influenced by the potentially large set of parameters that define the model. Optimizing a model means improving the match between the model's trajectory and the observed or desired temporal path at many individual points in time.

Recent advances in machine learning have greatly improved the potential to optimize differential equation models \autocite{chen18neural,bonnaffe21neural,rackauckas20universal}. However, with actual data, it is often not so easy to realize the promise of the new conceptual advances and software packages.

This article illustrates how to fit ordinary differential equation (ODE) models to noisy time series data. The fitted models are also sampled to develop an approximate Bayesian posterior distribution of trajectories. The posterior distribution provides a way to evaluate confidence in the fit to observed data and in the prediction of future observations.

I use the classic data for the fluctuations of lynx and hare populations, an example of predator-prey dynamics \autocite{odum71fundamentals}. This example illustrates the challenges that arise when fitting models for any scientific problem that can be analyzed by simple deterministic ODEs.

My work follows on the excellent recent article by Bonnaffé et al. \autocite{bonnaffe21neural} They fit neural ODEs (NODEs) to the hare-lynx data. NODEs use neural networks to fit time series data to the temporal derivatives of variables, in other words, NODEs estimate ODEs by using modern neural networks \autocite{chen18neural}. Bonnaffé et al.\autocite{bonnaffe21neural} emphasized that NODEs have the potential to advance many studies of ecological and evolutionary dynamics. However, they encountered several practical challenges in their application of NODEs to the hare-lynx data, ultimately concluding that ``it is our view that the training of these models remains nonetheless intensive.''

\section{Materials and Methods}

Extending Bonnaffé et al.\autocite{bonnaffe21neural}, I advance practical aspects of fitting ODE and NODE models. I show that several computational techniques provide a relatively easy way to fit and interpret such models. The computer code provides the specific methods by which I achieve each advance. Here, I emphasize six points.

\subsection{Comparing NODE and ODE models}

First, my computer code provides a switch between fitting the same data to a high dimensional NODE or a simple low dimensional ODE. \textit{Dimensionality} here refers to the size of the parameter space. The switch makes it easy to compare the two types of model. 

Conceptually, both NODE and ODE models are basic systems of ordinary differential equations. In practice, the difference concerns the typically much greater dimensionality of the NODE models and the wide variety of high-quality tools available to build, compute, and evaluate complex neural network architectures. The NODE models usually have much greater flexibility and power to fit complex patterns but also suffer from computational complexity and a tendency to overfit data.

\subsection{Dummy variables}

Second, I evaluate the costs and benefits of adding dummy variables to the fitting process. For example, there are two variables in the case of hare and lynx. To those two variables, we may add additional variables to the system. We may think of those additional dummy variables as unobserved factors \autocite{bonnaffe21neural}. For example, if there is one additional factor that significantly influences the dynamics, then trying to fit a two-dimensional model to the data will be difficult because the actual trajectories trace pathways in three dimensions. \textit{Dimensionality} here refers to the number of variables in the system.

\subsection{Data smoothing}

\begin{figure}[t]
\centering
\includegraphics[width=\ifmulticol3.42in\else4in\fi]{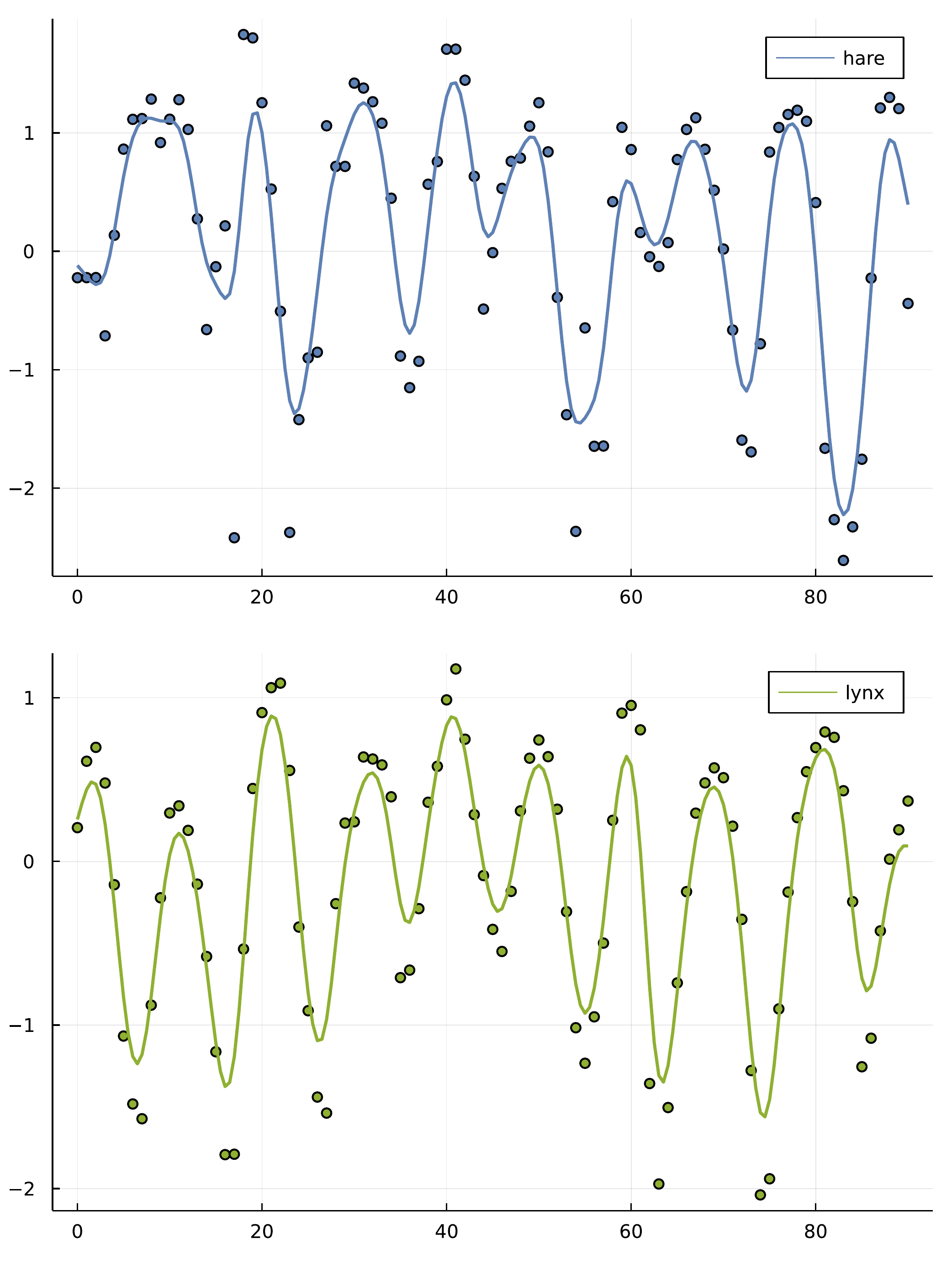}
\caption{The hare and lynx (circles) and smoothed (curves) data. The $x$-axis shows time in years, with one original observation per year over the 91 years of data collection. All models were fit to the smoothed curves at half-yearly intervals, creating 181 time points for each species. For each curve, the original data were first transformed by the natural logarithm and then normalized by the average value of the log-transformed observations. Data from Bonnaffé et al.\autocite{bonnaffe21neural}, originally from Odum \& Barrett \autocite{odum71fundamentals}.}
\label{fig:data}
\end{figure}

Third, I smoothed the original data before fitting the models. Deterministic models can only fit the general trends in the data. Stochastic fluctuations may interfere with the fitting process, which typically gains from a smoother cost function, in which the cost function decreases as the quality of the fit improves. To reduce large fluctuations, I first log-transformed the data and normalized by subtracting the mean for each variable of the times series \autocite{bonnaffe21neural}. Then, to emphasize trends in the data, I smoothed the time series with a cubic spline. I added an interpolated time point between each pair of observed time points. I then smoothed the augmented data with a Gaussian filter. \Figure{data} shows the original and the smoothed data.

\subsection{Sequential fitting}

Fourth, simultaneously fitting all points in the time series may prevent finding a good fit. The complexity of the optimization surface may be too great when starting from random parameters. Sequential fitting can help, first fitting the initial part of the time series, then adding later time points in a stepwise manner. However, when adding additional points, weighted equally with prior points, a strong discontinuity may arise in the fitting process. That discontinuity may push the fitting process too far away to recover a smooth approach to a good fit. To fix that issue, I slowly increased the weighting of later points, which provided greater continuity in the optimization process and better convergence to relatively good fits.

\subsection{Approximate Bayesian posterior}

Fifth, I estimate a distribution of temporal trajectories for a fitted model by analogy with sampling the Bayesian posterior distribution of the model parameters. The distribution of trajectories provides a measure of the confidence in the quality of a fit to the data and of predictions for future unobserved observations.

To sample the posterior distribution of fitted parameters and associated trajectories, I first fit the model by standard gradient descent methods, using the Adam learning algorithm \autocite{kingma14adam:}. Then, with the fitted model as an initial condition, I calculated the preconditioned stochastic gradient Langevin dynamics (pSGLD) \autocite{li15preconditioned}. In essence, a deterministic force moves each parameter toward a locally better fit, and a stochastic force causes parameter fluctuations.

For a gradient of the loss function with respect to the parameter, $g$, and a given hyperparameter, $\Ge$, the stochastic force dominates the deterministic force when $\Ge g \ll 1$. Thus, when the fit is sufficiently near a local optimum and the associated gradient multiplied by $\Ge$ is small, the model parameter value fluctuates randomly in a way that approximately samples the Bayesian posterior. Here, \textit{hyperparameter} means a parameter that controls the fitting process rather than a fitted parameter of the model, following the common convention in the machine learning literature \autocite{goodfellow16deep}.

In this study, I used a standard machine learning quadratic loss function \autocite{goodfellow16deep}. In particular, the loss is the sum of squared deviations between each target time point in the smoothed data and the value of the model trajectory at that time point.

A quadratic loss associates with the Bayesian estimator for the mean of a parameter's posterior distribution. However, in this study, I used pSGLD to sample the distribution of parameters around a local optimum for a fitted model, in which each observation is a multidimensional parameter vector describing a differential equation model. That distribution provides a rough estimate of the confidence in the fitted parameter values and the associated trajectories, inspired by Bayesian principles rather than adhering strictly to the assumptions of Bayesian analysis.

\subsection{Julia computer language}

Finally, I used the Julia programming language \autocite{bezanson17julia:}. Debate about choice of language often devolves into subjective factors. However, in my interpretation for fitting differential equation models, the current status of Julia provides several clear advantages.

Julia is much faster than popular alternatives, such as Python and R \autocite{perkel19julia:}. Speed matters, transforming difficult or essentially undoable optimizations into problems that can easily be solved on a standard desktop computer. I did all of the runs for this article on my daily working desktop computer using only the CPU, completing the most complex runs in at most a few hours, without any special effort to optimize the code or the process. Many useful runs for complex fits could be done in much less than an hour.

The Julia package DifferentialEquations.jl has a very wide array of numerical solvers for differential equations \autocite{rackauckas17differentialequations.jl--a}. Using an appropriate solver with the correct tolerances is essential for fitting differential equations. I used the solvers Rodas4P for ODE problems and TRBDF2 for NODE problems. These solvers handle the instabilities that frequently arise when fitting oscillatory time series data. Further experimentation would be useful to test whether other solvers might be faster or handle instabilities better.

Efficient optimization of large models typically gains greatly from automatic differentiation. \autocite{baydin18automatic,margossian19a-review} In essence, the computer code automatically analyzes the exact derivatives of the loss function with respect to each parameter, rapidly calculating the full gradient that allows the optimization process to move steadily in the direction that improves the fit.

For fitting differential equations, the special challenge arises because each time point along the target trajectory to be fit must be matched by using the differential equation solver to transform the model parameters into a predicted time point along the calculated trajectory. That means that differentiating the loss function with respect to the parameters must differentiate through the numerical solver for the system of differential equations. It must do so for each target point. In this study, the target trajectory consisted of 362 time points, requiring each calculation of a loss function or derivative of the loss to analyze the match between the data and 362 numerically evaluated trajectory points.

The Julia package DiffEqFlux.jl provides automatic differentiation through many different solvers \autocite{rackauckas20universal}. Other languages provide similar automatic differentiation but, in my experience, the process is either much slower or more limited in a variety of ways. By contrast, the Julia package works simply and quickly, with many options to adjust the process.

The DiffEqFlux.jl package also provides a broad set of tools to build neural network models. Those models can easily be analyzed as systems that estimate differential equations, NODEs which can be optimized with a few lines of code \autocite{rackauckas19diffeqflux.jl-a}.

Documentation of the DiffEqFlux.jl package presents several examples of fitting differential equation models. However, the toy data sets do not bring out many of the challenges one faces when trying to fit the kinds of noisy data that commonly arise in practice. The methods discussed here may be broadly useful for many applications.

\subsection{Overview of the models}

Each model has $n$ variables, two for hare and lynx and $n-2$ for dummy variables tracking unobserved factors. The models seek to match the log-transformed and smoothed data shown in \Fig{data}. The differential equation for the vector of variables $\bu$ in the ODE models has the form
\begin{equation}\label{eq:odede}
  \dovr{\bu}{t} = f\lr{\bS\bu - \bb},
\end{equation}
in which the $n^2+n$ parameters are in the $n\times n$ matrix, $\bS$, and the $n$ vector, $\bb$. The function $f$ maps the $n$ dimensional input to an $n$ dimensional output, potentially inducing nonlinearity in the model. One typically selects $f$ from the set of common activation functions used in neural network models. For all runs reported here, I used $f=\tanh$ applied independently to each dimension. It would be easy to study alternative ODE forms. However, in this article, I focus on \Eq{odede}.

For each numerical calculation of a predicted trajectory, I set the initial condition to the data values for the first two dimensions. For the other $n-2$ dimensions, I set a random initial value at the start of an optimization run. My code includes the option to optimize the initial values for the $n-2$ dummy variables by considering those values as parameters of the model. However, in my preliminary studies, optimizing the dummy initial conditions did not provide sufficient advantages. I did not use that option in the runs reported here.

Typical NODE models are neural networks that take $n$ inputs \autocite{chen18neural}. The $n$ outputs are the vector of derivatives, $\dd\bu/\dd t$. Common neural network architectures provide a variety of systems for calculating outputs from inputs. The Julia software packages include simple ways to specify common and custom architectures.

I used a simple two-layer architecture, in which each of the $n$ inputs flows to $N$ internal nodes. Each internal node produces an output that is a weighted sum of the inputs, each weight a model parameter, plus an additional parameter added as a constant. Each of those outputs is transformed by an activation function, $f$, for which I again chose $f=\tanh$. Those $N$ outputs were then used as inputs into a second layer that produced $n$ outputs as the weighted sum of inputs plus a constant. The second layer did not use an activation function to transform values.

This basic two-layer architecture has $(n+1)N+n(N+1)$ parameters. The runs in the following section used $N=20$ and $n=2,3,4$, yielding 102, 143, and 184 parameters for the increasing values of $n$.

The parameters and output for all computer models are included with the source code files.

\begin{figure*}[tbp]
\centering
\includegraphics[height=0.93\vsize]{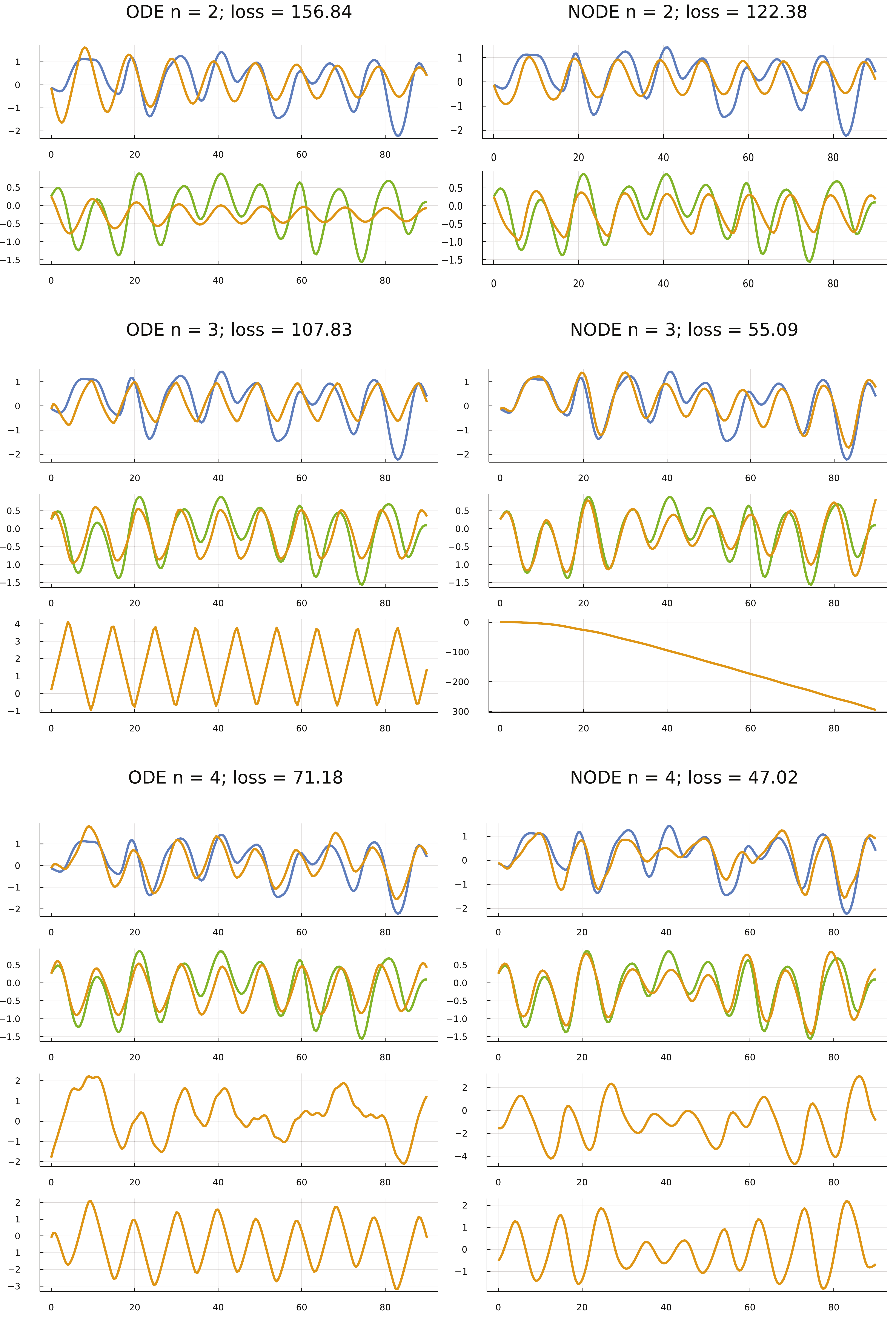}
\caption{Fit of models to the smoothed data. Gold curves show predicted trajectories. In each set, the top plot shows hare data (blue), and the second plot shows lynx data (green). Subsequent plots show dummy variables when $n>2$. The loss sums the squared deviations between the smoothed data and the model trajectories. I measured the deviations for both species at the 181 half-yearly intervals, yielding 362 squared-deviation components in the loss calculations.}
\label{fig:dynamics}
\end{figure*}

\begin{figure*}[tbp]
\centering
\includegraphics[width=\hsize]{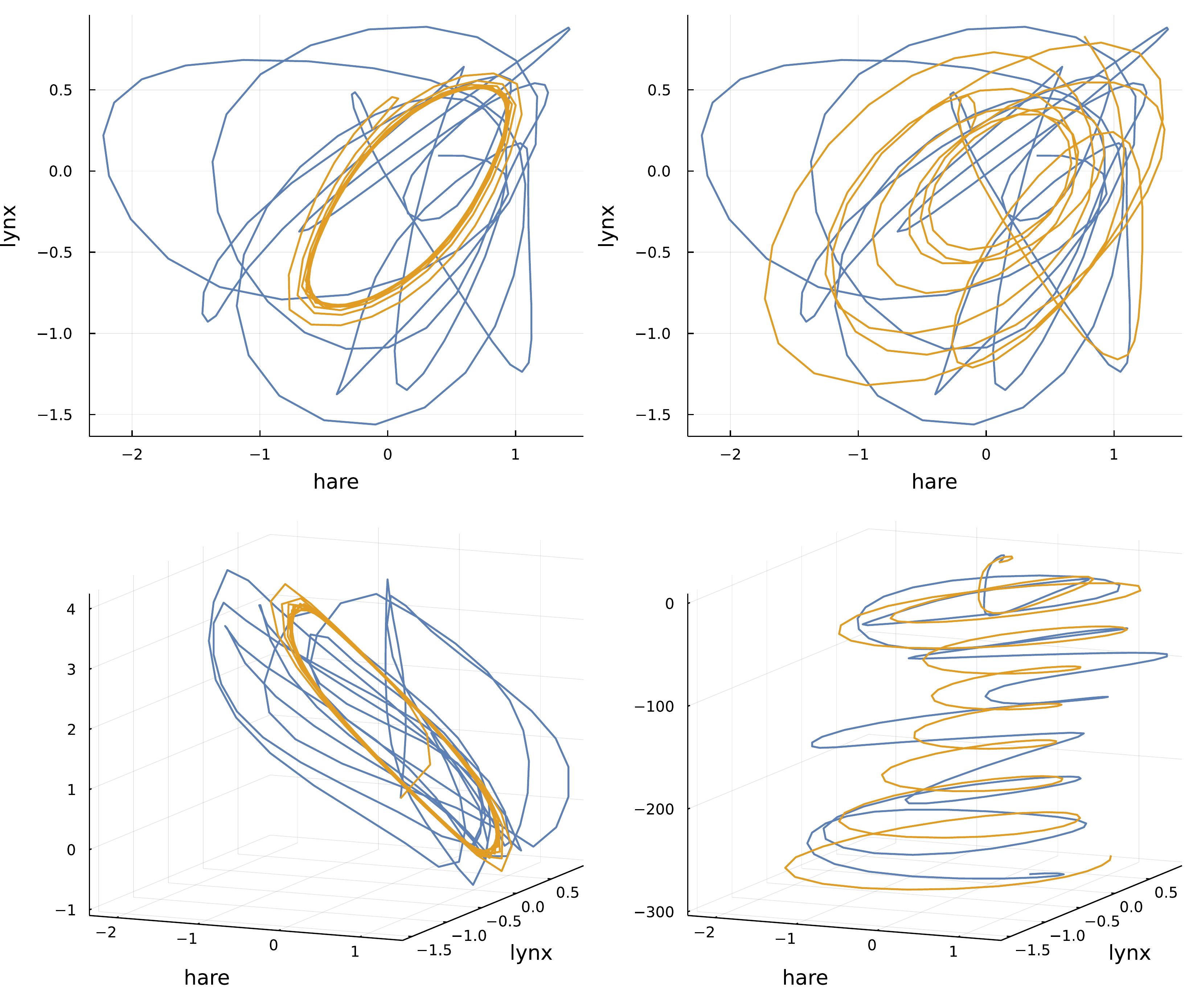}
\caption{Phase plots for ODE (left) and NODE (right) models with $n=3$. Blue curves trace the data trajectories, and gold curves trace the models' predicted trajectories. The upper plots show only the hare and lynx variables for the data and model, tracing $n=2$ dimensional trajectories. The lower plots add the third dummy dimension variable from the model to both the data and model trajectories.}
\label{fig:phase}
\end{figure*}

\begin{figure*}[tbp]
\centering
\includegraphics[height=0.85\vsize]{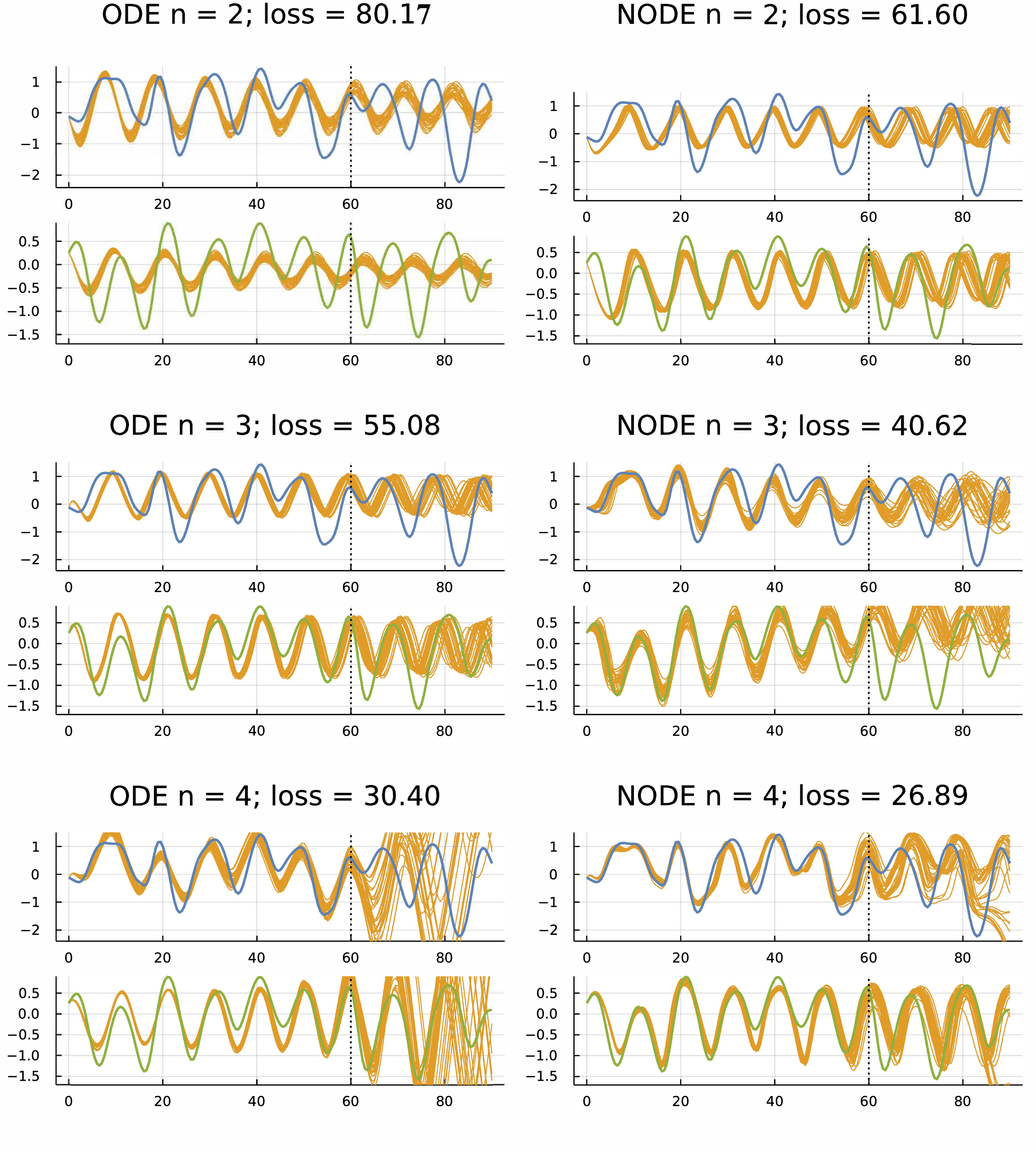}
\caption{Fit of models to the first 61 yearly observations (0--60) of the 91 yearly observations (0--90). The vertical dotted line shows the end of the fitted training period. The subsequent 30 years from 60 to 90 comprise the predictions of the model relative to data not used in fitting. The layout of the plots and colors for the various curves match \Fig{dynamics}, with blue for smoothed hare data, green for smoothed lynx data, and gold for the predicted trajectories of the models. For the predicted trajectories, I randomly chose 30 parameter combinations from the posterior distribution of parameters obtained by pSGLD sampling. The loss values are calculated for the training period, 0--60.}
\label{fig:bayes}
\end{figure*}

\section{Results}

\subsection{ODE versus NODE, varying $n$}

I fit ODE and NODE models for $n=2,3,4$. For each of those six combinations, \Fig{dynamics} shows the model fits against the smoothed data for the full 90-year period of observations.

The hare and lynx estimated abundances comprise the $n=2$ core variables of the analysis. Adding dummy variables to make $n>2$ improves the fit. That improving fit with rising $n$ can be seen in \Fig{dynamics} by the better match between the data and model trajectories as one moves down the fitting sets in each column.

NODE models fit better than ODE models, associated with the greater number of parameters in NODE models. The better fits can be seen by comparing the NODE sets in the right column against their matching ODE sets in the left column.

One expects better fits by adding variable dimensions to increase $n$ or adding parameter dimensions in NODE models. The benefit here is to see exactly how the fits change with the different changes in the models. For example, the phase plots in the next subsection show clearly how the constraints of the relatively low-dimension parameter space of the ODE models limits the fit when compared to the flexibility of the larger parameter space in the NODE models.

\subsection{Phase plots}

\Figure{dynamics} shows temporal trajectories for each species and each dummy variable, plotting abundance versus time. By contrast, phase plots draw trajectories by mapping the $n$ variables at each time to a point in $n$-dimensional space. Combining the $n$-dimensional points at different times traces a trajectory through phase space.

\Figure{phase} shows phase plots for ODE and NODE models with $n=3$ variables. The upper plots limit the trajectories to the $n=2$ dimensions for the hare and lynx data (blue) and model predictions (gold). In two dimensions, the trajectories do not match well. However, one can see that the ODE model in the upper left traces a regular cycle confined to a small part of the two-dimensional phase space, whereas the NODE model in the upper right moves widely over the space.

The three-dimensional phase plots in the lower panels of \Fig{phase} clarify the differences between the ODE and NODE models. For those three-dimensional plots, I used the third dimension from the models' dummy variable to augment both the data and the model prediction trajectories. For the ODE model in the lower-left panel, the model's phase trajectory remains confined to a limited part of a two-dimensional plane, whereas the data wander over the third dimension.

Adding the third dummy variable for the NODE model in the lower-right panel greatly enhances the match between the model predictions and the data. One can see the trajectory of that third variable in \Fig{dynamics}, right column, middle set for NODE and $n=3$, in the bottom panel of that set. In that case, the dummy variable starts near an abundance of 1 $(e^0)$ and declines toward 0 $(e^{-300})$.

By spreading the two-dimensional hare and lynx dynamics over a third dimension that declines steadily with time, the messy and visually mismatched data and model trajectories in the two-dimensional phase plot shown in the upper-right panel of \Fig{phase} are transformed into smoothly oscillating and matching three-dimensional trajectories in the lower-right panel of that figure.

It could be that the high-dimensional and flexible parameter space of the $n=3$ NODE model has discovered the simple geometry of the phase dynamics. One could of course find that geometry in other ways. The main advantage of the NODE model is that it does the fit quickly and automatically without any explicit assumptions about the shape of the dynamics.

\subsection{Predicting future observations}

Which models do best at predicting future outcomes? Given a single time series, one typically addresses that question by splitting the data. The first training subset provides data to fit the models. The second test subset measures the quality of the predictions.

I trained the six models in \Fig{dynamics} on the data from the first 61 yearly observations (0--60). I then compared the predictions of those models to the observed data for the subsequent 30 years (60--90). \Figure{bayes} shows the results.

I obtained predicted values for a model by calculating the model's temporal trajectory for a particular set of fitted parameters. The predictions are the temporal trajectory over the test period, the years 60--90. A single trajectory represents the predictions for one set of fitted parameters.

When making predictions, one wants an estimate of the predicted values and also a measure of confidence in the predictions. How much variability is there in the trajectories when using alternative sets of fitted parameters?

To obtain a distribution of fitted parameter sets, I used the pSGLD method described earlier. That method provides a Bayesian-motivated notion of the posterior parameter distribution. To draw the predicted gold trajectories in \Fig{bayes}, I estimated the posterior parameter distribution and then randomly sampled 30 parameter sets from that distribution.

How do the different models in \Fig{bayes} compare with regard to the quality of their predictions during the test period, 60--90? Starting at the top left, ODE $n=2$, the model predictions are precise but inaccurate. The small variation in trajectories during the test period reflects the high precision, whereas the large differences between the predictions and the data with regard to the timing of the oscillations reflect the low accuracy.

The low accuracy (high loss) during the training period and poor fit during the test period suggest that this model is underfit. Here, \textit{underfit} roughly means that the dimensionality of the variables or of the parameters is not sufficient to fit the data.

Next, consider the lower left model in that figure, ODE $n=4$. That model has high accuracy during the training period, but very low precision during the prediction period. The model seems to be overfit.

During the prediction period, the NODE models for $n=3,4$ also have relatively low precision and varying but typically not very good accuracy. Those models may also be overfit, for which \textit{overfit} means roughly that the models' high dimensionality caused such a close fit to the fluctuations in the training data that the models failed to capture the general trend in the data sufficiently to predict the outcome in the test period.

Finally, consider the two best models with regard to predictions during the test period, ODE $n=3$ and NODE $n=2$. Those models have intermediate accuracy during the fitted period, which seemed to avoid underfitting and overfitting. During the test period, both models had moderately good accuracy with regard to the timing of oscillations and moderately good precision with regard to variation in the predicted trajectories. Although the fits are far from perfect, they are good given the short training period in relation to the complex shape of the dynamics.

A technical challenge arises when deciding how long to run the sampling period for pSGLD and what hyperparameters to use to control that process. When does one have a sufficient estimate for the posterior distribution of trajectories? In a typical run for this study, I first ran the pSGLD sampling for a warmup period that created 5,000 parameter sets and associated trajectories. I collected 10,000 or more parameter sets and associated trajectories by pSGLD. For each trajectory, I calculated the loss for the full time over both the training and test periods.

I concluded that the sample was sufficient when the distribution of loss values for the first half of the generated parameter sets was reasonably close to the distribution of loss values for the second half of the generated parameter sets. As long as the loss distributions were not broadly different, the plotted trajectory distributions typically did not look very different.

One could also study posterior distributions for individual parameters. However, in this study, there was no reason to analyze individual parameters.

\section{Discussion}

The various technical advances greatly enhance the ease of fitting alternative differential equation models. New possibilities arise to analyze dynamics, gain insight into process, improve predictions, and enhance control.

In this article, I focused on fitting observed dynamics from a natural system. Alternatively, one could study how to design a system to achieve desired dynamics or to match a theoretical target pattern \autocite{hiscock19adapting}.

The pSGLD method to sample parameter combinations near a local optimum also raises interesting possibilities for future study. Technically, it is a remarkably simple and computationally fast method. Conceptually, it creates a kind of random walk near a local optimum on a performance surface, similar to mutation-selection dynamics near a local optimum of a fitness landscape \autocite{neher11statistical}. That analogy suggests the potential to gain further understanding of genetic variation and evolutionary dynamics on complex fitness surfaces.

\section*{Acknowledgments}

\noindent The Donald Bren Foundation, National Science Foundation grant DEB-1939423, and DoD grant W911NF2010227 support my research.

\section{Data availability}

All data and code are available at \url{https://github.com/evolbio/FitODE} and on Zenodo at \url{https://doi.org/10.5281/zenodo.6463624}. The parameters and output used to generate the figures in this article are only available at Zenodo.


\mybiblio	

\addcontentsline{toc}{section}{Appendix}


\end{document}